\documentclass[useAMS,usenatbib,usegraphicx]{mn2e}
\title[Evolution of Dust Temperature of Galaxies]
{Evolution of Dust Temperature of Galaxies through Cosmic Time as seen by {\it Herschel}
\thanks{Herschel is an ESA space observatory with science instruments provided
  by European-led Principal Investigator consortia and with important participation from NASA.}}
\author[H.S.~Hwang et al.]
{\parbox{\textwidth}{H.S.~Hwang,$^{1}$\thanks{E-mail: \texttt{hoseong.hwang@cea.fr}}
D.~Elbaz,$^{1}$
G.~Magdis,$^{1}$
E.~Daddi,$^{1}$
M.~Symeonidis,$^{2}$
B.~Altieri,$^{3}$
A.~Amblard,$^{4}$
P.~Andreani,$^{5,6}$
V.~Arumugam,$^{7}$
R.~Auld,$^{8}$
% ...
}\vspace{0.4cm}\\
\parbox{\textwidth}{$^{1}$Laboratoire AIM-Paris-Saclay, CEA/DSM/Irfu - CNRS - Universit\'e Paris Diderot, CE-Saclay, pt courrier 131, F-91191 Gif-sur-Yvette, France\\
$^{2}$Mullard Space Science Laboratory, University College London, Holmbury St. Mary, Dorking, Surrey RH5 6NT, UK\\
$^{3}$Herschel Science Centre, European Space Astronomy Centre, Villanueva de la Ca\~nada, 28691 Madrid, Spain\\
$^{4}$Dept. of Physics \& Astronomy, University of California, Irvine, CA 92697, USA\\
$^{5}$ESO, Karl-Schwarzschild-Str. 2, 85748 Garching bei M\"unchen, Germany\\
$^{6}$INAF - Osservatorio Astronomico di Trieste, via Tiepolo 11, 34143 Trieste, Italy\\
$^{7}$Institute for Astronomy, University of Edinburgh, Royal Observatory, Blackford Hill, Edinburgh EH9 3HJ, UK\\
$^{8}$Cardiff School of Physics and Astronomy, Cardiff University, Queens Buildings, The Parade, Cardiff CF24 3AA, UK\\
% ...
}}

\begin{document}

\date{Accepted 1988 December 15. Received 1988 December 14;
in original form 1988 October 11}

\pagerange{\pageref{firstpage}--\pageref{lastpage}} \pubyear{2002}

\maketitle

\label{firstpage}

\begin{abstract}
We study the dust properties of galaxies in the redshift range $0.1\la z\la 2.8$
  observed by the {\it Herschel Space Observatory}
  in the field of the Great Observatories Origins Deep Survey-North as part of PEP and HerMES key programmes.
Infrared (IR) luminosity ($L_{\rm IR}$) and dust temperature ($T_{\rm dust}$) of galaxies 
  are derived from the spectral energy distribution (SED) fit
  of the far-infrared (FIR) flux densities obtained with PACS and SPIRE instruments onboard {\it Herschel}.
As a reference sample, we also obtain IR luminosities and dust temperatures of local galaxies 
  at $z<0.1$ using {\it AKARI} and {\it IRAS} data in the field of the Sloan Digital Sky Survey.
We compare the $L_{\rm IR}-T_{\rm dust}$ relation between the two samples and find that: 
  the median $T_{\rm dust}$ of {\it Herschel}-selected galaxies at $z\ga$0.5 
  with $L_{\rm IR}\ga$5$\times10^{10}{\rm L}_\odot$, appears to be
  2--5\, K colder than that of 
  {\it AKARI}-selected local galaxies with similar luminosities; and
  the dispersion in $T_{\rm dust}$ for high-$z$ galaxies increases with $L_{\rm IR}$
  due to the existence of cold galaxies that are not seen among local galaxies.
We show that this large dispersion of the $L_{\rm IR}-T_{\rm dust}$ relation can 
  bridge the gap between local star-forming galaxies and high-$z$ submillimeter galaxies (SMGs).
We also find that three SMGs with very low $T_{\rm dust}$ ($\la20$ K) covered in this study
  have close neighbouring sources with similar 24-$\mu$m brightness,
  which could lead to an overestimation of FIR/(sub)millimeter fluxes of the SMGs.
 %  which suggests that $T_{\rm dust}$ measurements 
%  might be affected by an overestimation of FIR/(sub)millimeter fluxes, or
%  that galaxies in interacting pairs might be colder 
 % as a signature of an earlier merger stage.
\end{abstract}


\section{Introduction}

Understanding star formation (SF) mechanisms in galaxies and 
  how star formation rate (SFR) of galaxies
  evolves through cosmic time are 
  key issues in the study of galaxy formation and evolution.
The SFR is closely related to dust properties such as temperature ($T_{\rm dust}$), 
  mass, opacity, emissivity, and spatial extent. %\citep{cal00}.
In order to investigate dust properties of galaxies,
   it is important to obtain data covering the complete infrared (IR) wavelength range
   that includes both the ``Wien'' and ``Rayleigh-Jeans'' sides of the peak
   of the IR spectral energy distribution (SED).  
Thanks to the advent of the {\it Herschel Space Observatory} \citep{pil10} 
  with its very wide wavelength coverage ($70-500\mu$m),
  we are now able to study  complete IR SEDs of high-$z$ galaxies at $z\sim1-3$.
%By comparing the dust properties of high-$z$ galaxies with local ones 
%  we can give a strong constraint on the models of 
%  evolution of SFA in galaxies through the cosmic time.

The ``Rayleigh-Jeans'' side of the IR SED of a galaxy 
  is crucial for the study of dust properties,
  but up to now available photometric data for high-$z$ galaxies have been limited to a
  small number of wavelength windows and to small regions in the sky.
Despite the modest wavelength coverage of IR SEDs for high-$z$ galaxies,
  it was suggested that luminous infrared galaxies 
  (LIRGs; $10^{11}{\rm L}_\odot <L_{\rm IR}<10^{12}{\rm L}_\odot$) 
  and ultraluminous infrared galaxies (ULIRGs; $10^{12}{\rm L}_\odot <L_{\rm IR}<10^{13} {\rm L}_\odot$) 
  at high redshifts may be colder 
  than their local counterparts (e.g., \citealt{rr04,rr05,saj06,sym09,sey10}; see also \citealt{muz10}).
For example, \citet{sym09} found that the IR SEDs of high-$z$ galaxies,
   on average, are peaked at longer wavelengths than those of local galaxies
  by comparing {\it Spitzer} 70-$\mu$m-selected galaxies at $0.1\leq z<2$
  with {\it IRAS} 60-$\mu$m-selected galaxies at $z<0.1$.
They also found that the peak of IR SEDs for local galaxies is shifted 
  to the shorter wavelengths as IR luminosity ($L_{\rm IR}$) increases,
  while the peak of IR SEDs for high-$z$ galaxies is located at a wider wavelength range 
 compared to local galaxies and varies little with $L_{\rm IR}$.
However, \citet{bmag09}, using the {\it Spitzer} data in the
 fields of Great Observatories Origins Deep Survey (GOODS; \citealt{dic03}) and 
 Far Infrared Deep Extragalactic Legacy survey (FIDEL),
% showed that IR SEDs of high-$z$ galaxies at $0.4<z<1.3$
% are not significantly different  from those of local galaxies (see also \citealt{chapin10}). %?And more?
 suggested that IR SEDs of high-$z$ galaxies at $0.4<z<1.3$
 are not significantly different from those of local galaxies.
Several studies have also been carried out which suggest that
  the apparent change with redshift is mainly a selection effect (e.g., \citealt{pope06,chapin10}). 

It certainly seems that at high redshifts there are much colder populations of galaxies
  than local galaxies with similar IR luminosities
  [e.g., (sub)millimeter galaxies (SMGs),
  \citealt{blain02sub,chap05,pope06,kov06,huynh07,cop08,cle08}]. 
The relation between IR luminosity and dust temperature has been used as a useful tool
  in order to understand the connection between galaxy populations such as (U)LIRGs and SMGs
  \citep{dun00,chap03,chap05,kov06,cha07,yang07,you09,dye09,chapin09,chapin10,cle10,gmag10lbg}.
Interestingly, in the $L_{\rm IR}-T_{\rm dust}$ plane,
  SMGs form a separate locus from the (U)LIRGs,
  which can suggest a different origin between the two.

In this paper,
  we investigate the IR SEDs and the dust properties of high-$z$ galaxies
  by taking advantage of the wide wavelength coverage ($70-500\mu$m)
  of the Photodetector Array Camera (PACS; \citealt{pog10}) and 
  Spectral and Photometric Imaging Receiver (SPIRE; \citealt{gri10}) instruments onboard {\it Herschel}.
In order to study how dust properties of high-$z$ galaxies are different from those of 
  their local counterparts,
  we construct a sample of galaxies at $z<0.1$ that were observed by
  the {\it AKARI} telescope \citep{mur07}.
The {\it AKARI} all-sky survey data contain flux density measurements up to 160 $\mu$m,
  which can probe the long-wavelength side of the peak of IR SEDs 
  of local galaxies in a way similar to that of {\it Herschel} for high-$z$ galaxies. 
Throughout, we adopt $h=0.7$ and a flat $\Lambda$CDM cosmology with density parameters 
  $\Omega_{\Lambda}=0.73$ and $\Omega_{\rm m}=0.27$.
  
%__________________________________________________________________

\section{Observations and Data}\label{data}

GOODS-North (hereafter GOODS-N) was observed by {\it Herschel}
  as part of the Guaranteed Time Key Programmes
    PACS Extragalactic Probe (PEP\footnote{http://www.mpe.mpg.de/ir/Research/PEP}) and
    {\it Herschel} Multi-tiered Extragalactic Survey (HerMES\footnote{http://hermes.sussex.ac.uk}; Oliver et al. 2010, in prep.).
Source extraction on these PACS and SPIRE images 
  was performed at the prior positions of {\it Spitzer} 24-$\mu$m-selected sources, 
  and details are described in \citet{berta10}, \citet{bmag10} and \citet{rose10}.
PACS measurements are above the confusion limit, and
  we used flux densities in PACS bands down to 3$\sigma$ limits of 
  3 and 5.7 mJy at 100 and 160 $\mu$m, respectively \citep{berta10}.
SPIRE measurements are used down to 5$\sigma$ limits of
  4.4, 4.8 and 7.6 mJy at 250, 350 and 500 $\mu$m, respectively. 
It is noted that these measurements lie below the 1-$\sigma$ SPIRE confusion limit of 
  5.8, 6.3, 6.8 mJy \citep{ngu10}. 
However, this limit is a spatially averaged statistical limit 
  which considers that galaxies are homogeneously distributed in the field and 
  all affected in the same way by close neighbours. 
In this study,
  we flag galaxies more ``isolated'' than others for which SPIRE flux densities 
  can potentially be more robust
  by using the higher spatial resolution 24 $\mu$m images 
  (to be explained in \S \ref{results}).  

By combining these catalogues with the existing multi-wavelength data, %(from X-ray to FIR),
  we made a band-merged catalogue of IR sources having flux density measurements 
  at {\it Spitzer} MIPS 24 and 70 $\mu$m, 
  {\it Herschel} PACS 100 and 160 $\mu$m, and
  {\it Herschel} SPIRE 250, 350 and 500 $\mu$m (see \citealt{elb10} for details).   
For 493 galaxies detected in at least one out of the five PACS/SPIRE bands,
  we computed the IR luminosity using the SED models of \citet[CE01]{ce01}
  by allowing renormalization of the templates.
The SED fit was applied to the flux densities at $\lambda_{\rm rest}\geq30\mu$m,
  and $L_{\rm IR}$ for all 493 galaxies was calculated.

In order to determine the dust temperatures,
  we fit the observational data with a modified black body (MBB) model by fixing
  the emissivity parameter to $\beta=1.5$ \citep{hil83,gor10}.
We used only the galaxies that satisfy the following conditions:
\begin{enumerate}
\item there should be at least one flux measurement shortwards to the FIR peak
  (i.e. $0.55\lambda_{\rm p}\leq\lambda_{\rm rest}<\lambda_{\rm p}$
  where $\lambda_{\rm p}$ is the rest-frame, peak wavelength of 
  the best-fit CE01 SED model and $\lambda_{\rm rest}$ is the rest-frame wavelength of the observed data);
\item there should be at least one flux measurement longwards to the FIR peak
 (i.e. $\lambda_{\rm p}\leq\lambda_{\rm rest}$); and
\item the FIR SED should be physical (convex, not concave),
  i.e. we reject galaxies with
      ($S_{100}$ or $S_{160}$) $\geq$ $S_{250}$ and $S_{250}\leq$ ($S_{350}$ or $S_{500}$), or
      ($S_{100}$ or $S_{160}$) $\geq$ $S_{350}$ and $S_{350}\leq S_{500}$.
\end{enumerate}

The above criteria ensure a well-sampled SED around the peak of the FIR emission,
  and the wavelength cut of $0.55\lambda_{\rm p}$ is introduced
  to reduce the contribution of warm dust (i.e. emission from very small grains).
%  and is determined through the experiment of MBB fit to model SED templates.
For these galaxies,
  we fit the flux densities at $\lambda_{\rm rest}\geq0.55\lambda_{\rm p}$
  using the MBB model.
We then select galaxies having flux measurements on both sides of 
  the peak for the best-fit MBB model, and 
  secure a final sample of 140 galaxies.
Uncertainties of $L_{\rm IR}$ and $T_{\rm dust}$ were computed by randomly selecting
  flux densities at each band within the associated error distribution (assumed to be Gaussian)
  and then re-fitting.
It is noted that we checked how $T_{\rm dust}$ measurements of galaxies
  are affected if we remove some data points for the fit.
We found that $T_{\rm dust}$ becomes systematically lower/higher
  if you use only data points shortwards/longwards to the FIR peak.
However, $T_{\rm dust}$ does not change systematically
  even if you remove some data points at $\lambda_{\rm rest}\geq0.55\lambda_{\rm p}$
  for the fit as long as you have measurements on both sides of the peak wavelength.

For local galaxies,
  we construct an IR catalogue by cross-correlating
  our IR sources with the galaxies in the redshift catalogue as follows. 
We use the {\it IRAS} Faint Sources Catalog -- Version 2 (\citealt{mos92}),
  which contains 173,044 sources
  with flux density measurements at 12, 25, 60 and 100 $\mu$m.
We also use the {\it AKARI}/Far-Infrared Surveyor (FIS; \citealt{kaw07})
  all-sky survey Bright Source Catalogue
  (BSC\footnote{http://www.ir.isas.jaxa.jp/AKARI/Observation/PSC/ 
  Public/RN/AKARI-FIS$\_$BSC$\_$V1$\_$RN.pdf}) ver. 1.0 
  that contains 427,071 sources over the whole sky,
  with  measured flux densities at 65, 90, 140 and 160$\mu$m.
Among the measurements at {\it IRAS} and {\it AKARI} bands,
  we consider only the reliable\footnote{Flux quality flags 
  are either `high' or `moderate' for {\it IRAS} sources
  and `high' for {\it AKARI} sources.} flux densities. 
For the redshift catalogue, we use a spectroscopic sample of galaxies in 
  the Sloan Digital Sky Survey Data Release 7 \citep[SDSS DR7]{aba09} 
  complemented by a photometric sample of galaxies
  whose redshift information is available in the literature \citep{hwa10}.

%Figure 1 %%%%%%%%%%%%%%%%%%%%%%%%%%%%%%
\begin{figure}
\center
\includegraphics[scale=0.45]{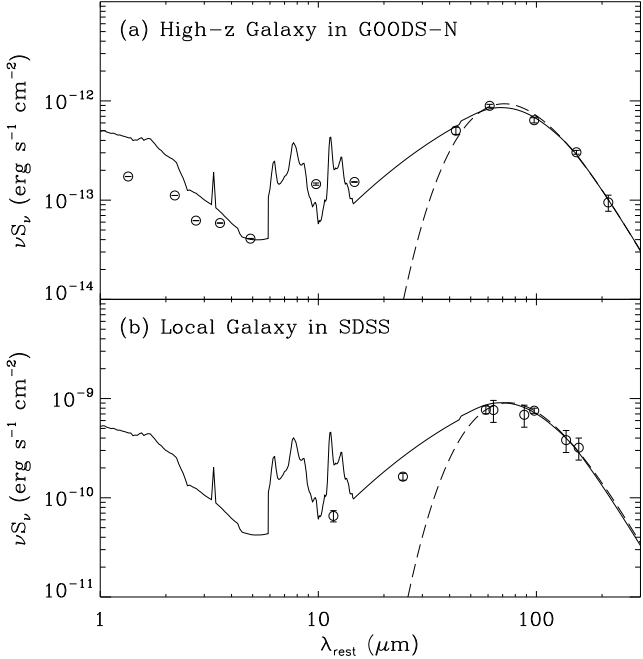}
\caption{Example SEDs for 
  (a) a high-$z$ galaxy in GOODS-N with 
  $L_{\rm IR}=6.7\times10^{11} ({\rm L}_\odot)$, $T_{\rm dust}=37$ K, and $z=0.64$;
  (b) a local galaxy in SDSS with 
  $L_{\rm IR}=4.9\times10^{11} ({\rm L}_\odot)$, $T_{\rm dust}=35$ K, and $z=0.02$.
The best-fit CE01 and MBB models are indicated by solid and dashed lines, respectively.
}\label{fig-fit}
\end{figure}
%%%%%%%%%%%%%%%%%%%%%%%%%%%%%%%%%%%%%%%%

%Figure 2 %%%%%%%%%%%%%%%%%%%%%%%%%%%%%%
\begin{figure}
\center
\includegraphics[scale=0.45]{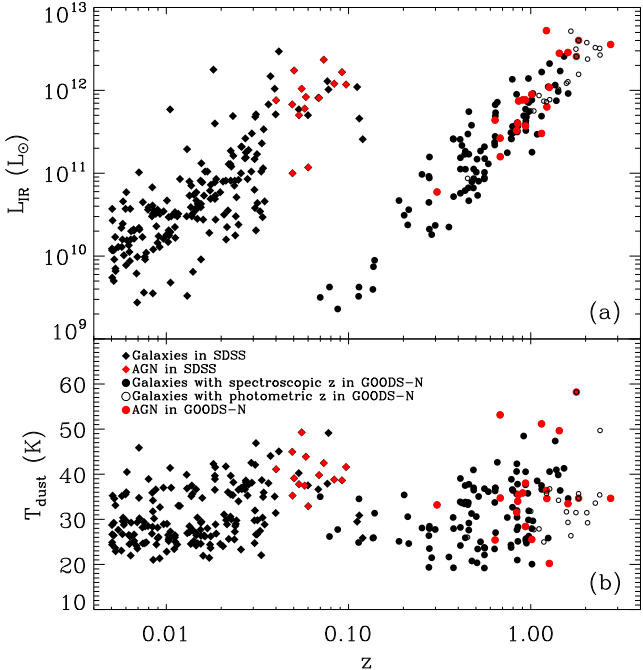}
\caption{(a) $L_{\rm IR}$ and (b) $T_{\rm dust}$
  for 140 galaxies in GOODS-N (circles) and 190 galaxies in SDSS (diamonds) 
  vs. redshift.
Galaxies with spectroscopic and photometric redshifts are
  filled and open circles, respectively. 
Galaxies hosting AGN are indicated by red symbols.
}\label{fig-z}
\end{figure}
%%%%%%%%%%%%%%%%%%%%%%%%%%%%%%%%%%%%%%%%

By adopting the same method applied to high-$z$ galaxies,
  we compute $L_{\rm IR}$ for the local galaxies 
  by fitting the CE01 templates to the flux densities at $\lambda_{\rm rest}\geq30\mu$m.
For $T_{\rm dust}$ estimates,
  in addition to the conditions (i) and (ii) used for high-$z$ galaxies,  
  we use only sources detected at 140 or 160 $\mu$m.
This condition ensures a fair comparison with the high-$z$ samples
  as in both samples have flux density measurements longwards to the FIR peak.
%We rejected galaxies that show a large discrepancy in flux densities 
%  between {\it IRAS} 60 $\mu$m ($S_{60}$) and {\it AKARI} 65 $\mu$m ($S_{65}$)
%  or between {\it IRAS} 100 $\mu$m ($S_{100}$) and {\it AKARI} 90 $\mu$m ($S_{90}$)
%  if both $S_{60}$ and $S_{65}$ (or $S_{100}$ and $S_{90}$) are measured 
 % (i.e. $\vert{\rm log}(S_{100}/S_{90})\vert>0.3$ or $\vert{\rm log}(S_{60}/S_{65})\vert>0.3$).
Since {\it IRAS} 60 $\mu$m and {\it AKARI} 65 $\mu$m 
  (or {\it IRAS} 100 $\mu$m and {\it AKARI} 90 $\mu$m) partially overlap,
  we use only {\it IRAS} data ($S_{60}$ and $S_{100}$) for the fit to avoid over-weighting
  when both {\it IRAS} and {\it AKARI} flux densities are measured.
Finally, we use only the galaxies having flux density measurements
  on both sides of the peak of the best-fit MBB model.

Fig. \ref{fig-fit} represents example SEDs for a high-$z$ galaxy in GOODS-N and a local galaxy in SDSS
  with the best-fits CE01 and MBB models.
In Fig. \ref{fig-z}, 
  we plot $L_{\rm IR}$ and $T_{\rm dust}$ for 140 high-$z$ and 190 local galaxies
   as a function of redshift,
   which shows that the two samples are distributed over similar ranges 
   of $L_{\rm IR}$ and $T_{\rm dust}$.
The high-$z$ galaxies in GOODS-N are found at $0.07<z<2.74$
  with $2.3\times10^{9} ({\rm L}_\odot)<L_{\rm IR}<5.3\times10^{12} ({\rm L}_\odot)$
  and $19$ (K)$<T_{\rm dust}<58$ (K),
  while the local galaxies in SDSS are in the range $0.005<z<0.119$
  with $2.7\times10^{9} ({\rm L}_\odot)<L_{\rm IR}<3.0\times10^{12} ({\rm L}_\odot)$
  and $21 $(K)$<T_{\rm dust}<49$ (K).
%This shows that the two samples are distributed in a similar range of $L_{\rm IR}$ and $T_{\rm dust}$.
%  which indicates that the observational selection effects are not significantly 
%  different between them.

%Figure 3 %%%%%%%%%%%%%%%%%%%%%%%%%%%%%%
\begin{figure*}
\center
\includegraphics[scale=0.75]{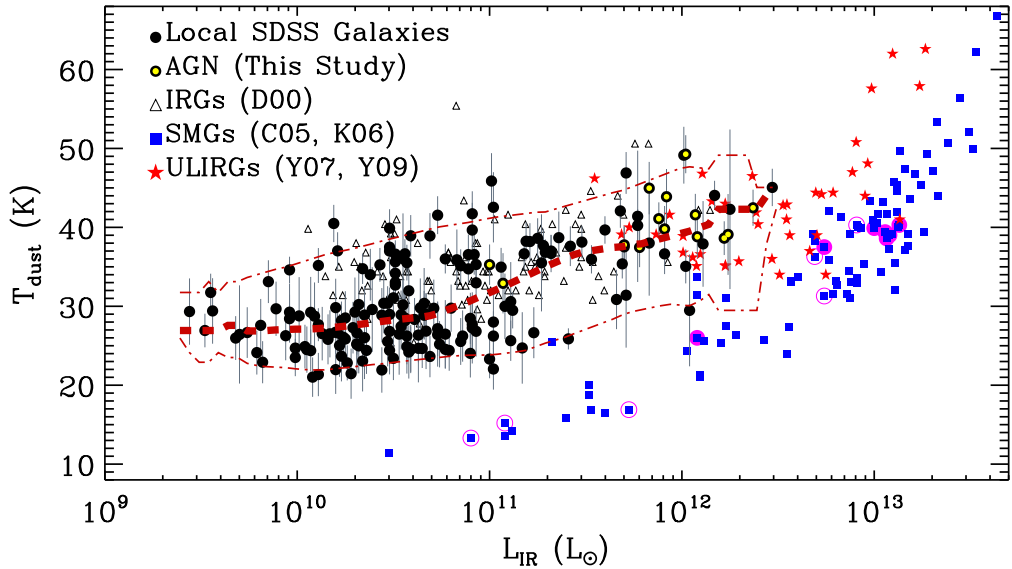}
\caption{$T_{\rm dust}$ vs. $L_{\rm IR}$ for galaxies in SDSS.
Galaxies hosting AGN are indicated by yellow symbols.
The thick dashed line is a smoothed median trend of $T_{\rm dust}$
  for local SDSS galaxies by excluding those with AGN,
  and the dot-dashed lines are its envelope that includes
  90\% of the galaxies above and below the median.  
The known local infrared galaxies (IRGs) (D00: \citealt{dun00}),
  SMGs (C05: \citealt{chap05}, K06: \citealt{kov06}) and 
  ULIRGs (Y07: \citealt{yang07}, Y09: \citealt{you09})
  are plotted with triangles, squares and star symbols, respectively
Among the SMGs in common between this study and C05,
   those having no neighbouring sources (clean) are denoted by large filled circles,
   while those possibly contaminated by neighbouring sources (blended) are denoted by large open circles.
}\label{fig-lir_tdlow}
\end{figure*}
%%%%%%%%%%%%%%%%%%%%%%%%%%%%%%%%%%%%%%%%

\section{Results}\label{results}

\begin{figure*}
\center
\includegraphics[scale=0.85]{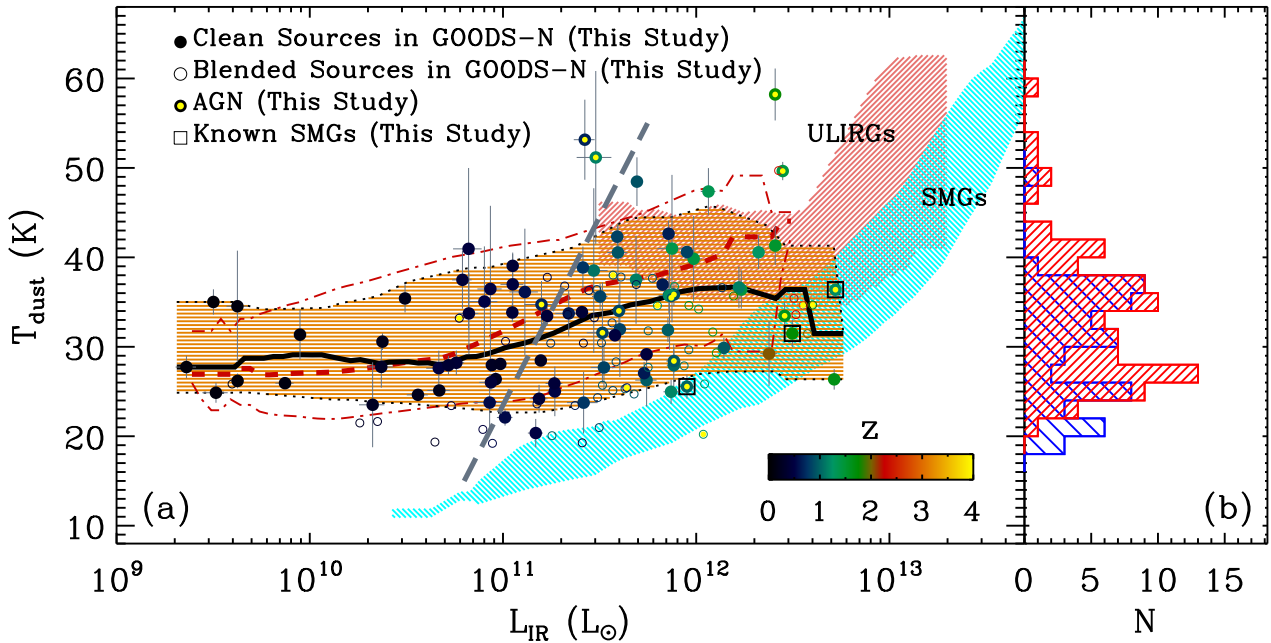}
\caption{(a) $T_{\rm dust}$ vs. $L_{\rm IR}$ for galaxies in GOODS-N.
Galaxies hosting AGN are indicated by yellow symbols.
The thick dashed line is a smoothed median trend of $T_{\rm dust}$
  for local galaxies adopted in Fig. \ref{fig-lir_tdlow},
  and the dot-dashed lines are its 90\% envelope. 
The loci of known SMGs and ULIRGs in Fig. \ref{fig-lir_tdlow} are plotted
  as regions filled by cyan and coral colour, respectively
Clean and blended galaxies are denoted by filled and open circles, respectively.
The thick solid line is a median trend of $T_{\rm dust}$ 
  for galaxies in the GOODS-N field and the dotted lines are its 90\% envelope 
  (filled with orange colour).
Inclined, grey long-dashed lines indicate
   the {\it AKARI} selection function at the maximum redshift of local galaxies ($z=0.119$).
(b) Distribution of $T_{\rm dust}$ for high-$z$ galaxies.
Clean and blended galaxies including AGN are denoted by hatched histograms with
  orientation of 45$^\circ$ ($//$ with red colour) and 
  of 315$^\circ$ ($\setminus\setminus$ with blue colour) relative to horizontal, respectively.
}\label{fig-lir_td}
\end{figure*}

We show the relation between $L_{\rm IR}$ and $T_{\rm dust}$ 
  for local and high-$z$ samples in Figs. \ref{fig-lir_tdlow} and \ref{fig-lir_td}, respectively.
Fig. \ref{fig-lir_tdlow} shows the distribution of local galaxies in comparison with
  known ULIRGs and SMGs in the literature.
We determine a smoothed median trend of $T_{\rm dust}$ as a function of $L_{\rm IR}$
  for our local galaxies.
Since the contribution of active galactic nuclei (AGN) to determining 
  the median trend of $T_{\rm dust}$ for local and high-$z$ galaxies could be different,
  we exclude them when determining the median.
%The median trend and its 2$\sigma$ scatter are plotted
 % to show how they distribute in the plane of $T_{\rm dust}$ and $L_{\rm IR}$.
We call AGN those sources whose optical spectral types are found to be Seyferts, 
  low-ionization nuclear emission-line regions (LINERs) or composite galaxies
  in the emission line ratio diagram \citep{bpt81,kew06}.
We classify only the galaxies at $z>0.04$ as AGN
  due to the problem of the small (3\arcsec) fixed-size aperture for SDSS spectroscopy \citep{kew06}.
It is seen that $T_{\rm dust}$ remains constant in the lower end of the luminosity range 
  ($L_{\rm IR}\la 5\times10^{10}{\rm L}_\odot$) and increases as $L_{\rm IR}$ increases.
This trend has been indicated in the previous studies 
  by a shift of the peak in the IR SED with increasing $L_{\rm IR}$
  (\citealt{soi87,soi89,chap03,chapin09,sym09}; see also \citealt{amb10}).
At $L_{\rm IR}\ga10^{12}$ (L$_\odot$),
  our samples are smoothly connected to the locus of known ULIRGs at intermediate/high redshifts.
We also note that changing AGN selection criteria has a small effect on our results.

%Figure 4 %%%%%%%%%%%%%%%%%%%%%%%%%%%%%%
%%%%%%%%%%%%%%%%%%%%%%%%%%%%%%%%%%%%%%%%

In Fig. \ref{fig-lir_td}(a), we plot high-$z$ galaxies
  along with the loci of ULIRGs, SMGs and local SDSS galaxies.
The median trend of $T_{\rm dust}$ %as a function of $L_{\rm IR}$ 
  for high-$z$ galaxies is also determined 
  by excluding those galaxies hosting X-ray selected AGN as well as possibly 
  blended sources.
AGN have either L$_{\rm X}$[0.5--8.0 keV] $>$ 3$\times$10$^{42}$ ergs s$^{-1}$, 
  a hardness ratio (ratio of the counts in the 2--8 keV to 0.5--2 keV passbands) 
  greater than 0.8, N$_{\rm H}\geq$10$^{22}$ cm$^{-2}$, 
  or broad/high-ionization AGN emission lines \citep{bau04}. 
We regard a galaxy as `blended' if it has at least one neighbouring source
  within $\sim$one beam FWHM of the galaxy position 
  at 24 $\mu$m and SPIRE images 
  ($20\arcsec$ for 24$\mu$m and 250 $\mu$m, $27\arcsec$ for 350 $\mu$m,
   and $46\arcsec$ for 500 $\mu$m). %from trial and error
For the 24-$\mu$m image, the criterion is at least two neighbouring sources
  instead of one.
In addition, we require that the flux density of a neighbouring source should be larger than 50$\%$ 
  of that of the galaxy.
Among 140 high-$z$ galaxies, we have 84 `clean' and 56 `blended' galaxies
  (see also \citealt{bris10} for a purity index that is a measure of blending of a galaxy).
Fig. \ref{fig-lir_td}(b) represents the distribution of $T_{\rm dust}$
  for `clean' and `blended' galaxies including AGN.
This shows that only `blended' galaxies have several cold galaxies with $T_{\rm dust}\sim20$ K,
  while `clean' galaxies do not (to be discussed in detail in \S \ref{cold}).

In Fig. \ref{fig-lir_td}(a), 
  the median trend of $T_{\rm dust}$
  for high-$z$ galaxies also 
  appears fairly constant at low luminosities, 
  but increases as $L_{\rm IR}$ increases at higher luminosities.
By comparing with local galaxies,
  we see that the median trend of $T_{\rm dust}$ 
  for {\it Herschel}-selected high-$z$ galaxies at $z\ga0.5$
  becomes smaller than that for
  {\it AKARI}-selected low-$z$ galaxies 
  by $2-5$ K at high luminosities ($L_{\rm IR}>5\times10^{10}{\rm L}_\odot$).
Note that this does not necessarily mean that high-$z$ galaxies are systematically
  colder than local galaxies with similar luminosities 
  (to be discussed in \S \ref{sel}).
It is also seen that  $T_{\rm dust}$ scatter increases by about 5 K
  in the sense that the upper envelope for high-$z$ galaxies
  is similar to that of local galaxies, 
  while the lower envelope for high-$z$ galaxies stretches below. 
These results are consistent with the results 
  based on the peak position of the IR SEDs \citep{sym09}.
Interestingly,
  the high-$z$ galaxies with low $T_{\rm dust}$ fill the gap 
  between local star-forming galaxies and high-$z$ SMGs 
  in the $L_{\rm IR}-T_{\rm dust}$ plane.
%This large dispersion of $T_{\rm dust}$ for high-$z$ galaxies 
%  leads that the median trend of $T_{\rm dust}$ for high-$z$ galaxies 
%  becomes smaller than that for
%  low-$z$ galaxies by $2-5$ K at high luminosities ($L_{\rm IR}>5\times10^{10}L_\odot$),
%This enlargement of the dispersion of high-$z$ galaxies leads
%  that the median trend of $T_{\rm dust}$ for high-$z$ galaxies
%  is smaller than that for low-z galaxies by $2-5$K
%  at high luminosity range ($L_{\rm IR}>5\times10^{10}L_\odot$),
%  and fills the gap between local star-forming galaxies and high-$z$ SMGs.
It is also interesting that
  $T_{\rm dust}$ for AGN follows a similar trend as for star-forming galaxies,
  which indicates that the star formation of galaxies having AGN does not significantly
  differ from that of normal star-forming galaxies 
  seen in the FIR regime, where SF dominates the IR emission of AGN \citep{elb10,shao10,hat10}.

%Figure 5 %%%%%%%%%%%%%%%%%%%%%%%%%%%%%%
\begin{figure}
\center
\includegraphics[scale=0.45]{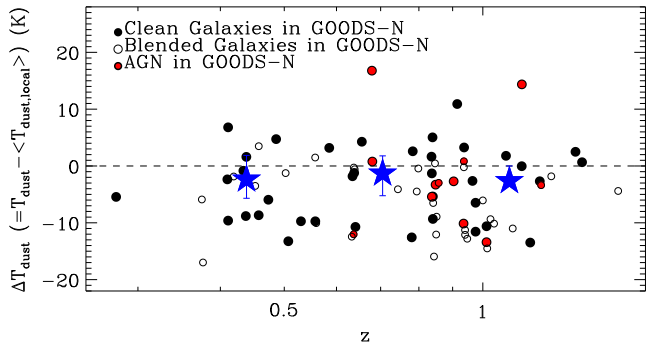}
\caption{Deviation of $T_{\rm dust}$ ($\Delta T_{\rm dust}$) for high-$z$ LIRGs 
  from the median trend of local LIRGs vs. redshift.
Filled and open symbols are `clean' and `blended' galaxies, respectively.
Star symbol represents the median value of $\Delta T_{\rm dust}$ 
  for `clean' galaxies in each redshift bin.
}\label{fig-tz}
\end{figure}
%%%%%%%%%%%%%%%%%%%%%%%%%%%%%%%%%%%%%%%%

To investigate how $T_{\rm dust}$ for high-$z$ galaxies evolves with redshift
  compared to local galaxies with similar $L_{\rm IR}$, in Fig. \ref{fig-tz},
  we plot the deviation of $T_{\rm dust}$ for high-$z$ LIRGs 
  from the median trend ($\langle T_{\rm dust,local}\rangle$, thick dashed line in Fig. \ref{fig-lir_td}a) 
  of $T_{\rm dust}$ for local LIRGs 
  ($\Delta T_{\rm dust}=T_{\rm dust}-\langle T_{\rm dust,local}\rangle$) as a function of redshift.
This shows that $\Delta T_{\rm dust}$ is negative, on average,
  which indicates that {\it Herschel}-selected high-$z$ LIRGs at $0.3<z<1.4$ appear
  to be colder than {\it AKARI}-selected local LIRGs.
$\Delta T_{\rm dust}$ changes little with redshift,
  but it is difficult to draw any strong conclusions due the small sample size.
Since the deeper observation of GOODS fields will be conducted
  in PEP and in the GOODS-{\it Herschel} key programme (PI: D. Elbaz),
  the future {\it Herschel} data will help us to address this issue
  with a larger number of galaxies up to higher redshift.

%Figure 6 %%%%%%%%%%%%%%%%%%%%%%%%%%%%%%
\begin{figure}
\center
\includegraphics[scale=0.43]{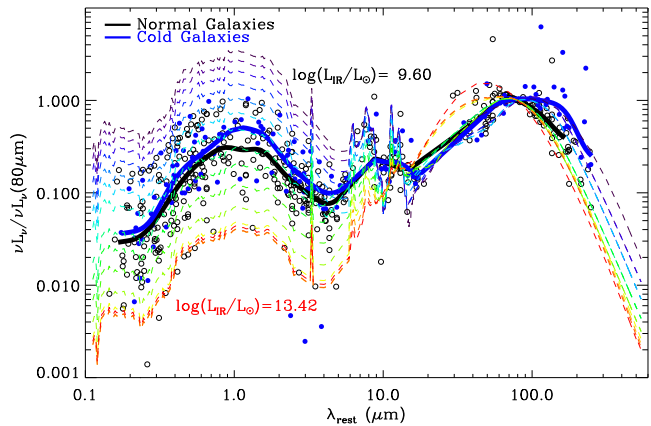}
\caption{Average SEDs of LIRGs in GOODS-N with spectroscopic redshifts
 normalized by the luminosity at 80 $\mu$m.
%  with $0.27\leq z\leq1.42$ and $11\leq {\rm log}(L_{\rm IR}/L_\odot)\leq12$.
Observed data for 10 {\it cold} and 26 {\it normal} galaxies
  are plotted by filled and open circles, respectively.
The average SEDs of the data points are plotted as solid lines.
Colour-coded dashed lines are CE01 templates
 normalized by the luminosity at 80 $\mu$m,
  and $L_{\rm IR}$ of the templates 
  with the maximum and the minimum $L_{\rm IR}$ are labelled.
}\label{fig-sed}
\end{figure}
%%%%%%%%%%%%%%%%%%%%%%%%%%%%%%%%%%%%%%%%

To investigate how the SEDs of high-$z$ galaxies
  are different from the local SED templates of CE01,
  we plot the average SEDs (from UV to FIR)
  of {\it cold} and {\it normal} LIRGs at high redshifts
  in comparison with the local SED templates of CE01 in Fig. \ref{fig-sed}.
The high-$z$ galaxies are classified into three groups
  by comparing their $T_{\rm dust}$ values with the median trend of $T_{\rm dust}$ for local galaxies 
  (thick dashed line in Fig. \ref{fig-lir_td}a)
: {\it cold} (those with $T_{\rm dust}$ values less than the lower envelope of local galaxies);
  {\it normal} (those with $T_{\rm dust}$ values within the envelopes of local galaxies); and
  {\it warm} (those with $T_{\rm dust}$ values larger than the upper envelope of local galaxies).
Fig. \ref{fig-sed} shows that the peak wavelength of the FIR SED for {\it cold} galaxies is larger
  than that for {\it normal} galaxies as expected.
The average FIR SED of  {\it normal} galaxies appears to match well with 
  the CE01 template of similar luminosity, but that of  {\it cold} galaxies 
  is matched with roughly 80 times less luminous template,
  even if the median $L_{\rm IR}$ of {\it normal} and {\it cold} galaxies are similar 
  ($\langle L_{\rm IR}/{\rm L}_\odot \rangle \sim5\times10^{11}$ and $4\times10^{11}$ ${\rm L}_\odot$
  for {\it normal} and {\it cold} galaxies, respectively).

We wish to emphasize that the longer-wavelength side of the peak of the FIR SED 
  for {\it cold} galaxies could be affected 
  by blending at long wavelengths.
Since the beam FWHM gets larger as wavelength increases,
  the flux densities at longer wavelength can be systematically overestimated
  if there is more than one source within the resolution limit.
In addition, galaxies having neighbouring sources seem to have low $T_{\rm dust}$
  compared to `clean' galaxies, as seen in Fig. \ref{fig-lir_td}(b),
  which suggests that we should be cautious
  when we measure $T_{\rm dust}$ using multi-wavelength data with different resolutions.
%This kind of contamination is relatively severer in cold galaxies
%  as shown in Fig. \ref{fig-lir_td}(c),
%  and we tried to eliminate them in our samples by identifying blended galaxies.
We have calculated the fraction of blended galaxies among
  {\it normal} and {\it cold} galaxies,
  and find values of 34$\pm6\%$ and 58$\pm9\%$, respectively. 
We have tried to eliminate such problem in our samples by identifying blended galaxies,
  and by using the FIR source catalogue based on prior extraction \citep{berta10,rose10}.
However, since we may not have completely deblend all sources and
  do not account for galaxies whose neighbours are
  less bright than 50$\%$ of the target galaxy,
  there could be a remaining contribution to the FIR flux of {\it cold} galaxies 
  due to this effect.
   
Interestingly, the {\it cold} galaxies are brighter than the {\it normal} galaxies
  in the optical bands.
This could imply that there is a correlation between $T_{\rm dust}$ and
  the optical properties such as galaxy morphology, colour, or size.
Thus, the study of the relation between optical properties and $T_{\rm dust}$ 
  may provide us useful hints for the origin of the {\it cold} galaxies 
  (Le Floc'h et al. 2010, in prep.).
  
\section{Discussion}\label{con}

\subsection{Selection Effects}\label{sel}
%In Fig. \ref{fig-lir_td}, we show that 
%  high-$z$ galaxies at $L_{\rm IR}>2\times10^{10}L_\odot$
%  seem to be colder than the local galaxies with similar $L_{\rm IR}$
%  by $2-5$K due to the existence of {\it cold} galaxies that are not seen in the local samples.

The {\it AKARI} data contain flux density measurements up to $\lambda=$ 160 $\mu$m
 ($\lambda_{\rm rest}=$ 157 $\mu$m with a median redshift of $z=0.015$ for local samples),
  while {\it Herschel} data have measurements up to $\lambda=$ 500 $\mu$m
 ($\lambda_{\rm rest}=$ 330 $\mu$m with a median redshift of $z=0.5$ for high-$z$ samples).
Therefore, the $T_{\rm dust}$ difference between the two samples seen in Fig. \ref{fig-lir_td}(a)
  could be affected by the difference in selection effects.
For example, \citet{chapin10} found results similar to this study,
  in the sense that the dust temperatures of {\it BLAST}-selected (250$-$500 $\mu$m)
  high-$z$ galaxies at $z<3$ are systematically cooler than 
  {\it IRAS} 60-$\mu$m-selected local galaxies.
When they account the difference in selection effects between the two samples,
  they find no evidence for evolution of $T_{\rm dust}$.
On the other hand, 
  if we consider {\it Herschel}-selected local galaxies ($z<0.1$) 
  observed in the {\it Herschel}-ATLAS key programme \citep{eal10},
  we can find cold galaxies with $T_{\rm dust}\la20$ K that are not
  seen in our local sample (see Fig. 2 in \citealt{amb10}).
This seems to support the idea that the different selection effects between 
  local and high-$z$ galaxies can explain the $T_{\rm dust}$ difference.  
However, note that there are few local galaxies with 
  $L_{\rm IR}\sim 10^{11}-10^{12} {\rm L}_\odot$ ($z<0.1$) in the sample of 
  \citet[see their Fig. 3]{amb10},
  which are crucial for understanding the $T_{\rm dust}$ difference 
  between local and high-$z$ galaxies (see Fig. \ref{fig-lir_td}a).
Since a wider area will eventually be covered by the {\it Herschel}-ATLAS program,
  the upcoming data will help us to address this issue.

To investigate the effect of the different selection effect 
  (i.e. observed wavelength and detection limit)
  on the difference of $T_{\rm dust}$ between local and high-$z$ galaxies,
  and on the existence of cold galaxies 
  ($T_{\rm dust}\la27$ K and $L_{\rm IR}> 3\times10^{11} {\rm L}_\odot$)
  that are not seen in the local sample but seen in the high-$z$ sample,  
  we made a following experiment:
  we move the best-fit SEDs of high-$z$ galaxies to local universe,
  and check whether they can be observed with {\it AKARI} detection limit 
    used for local galaxies in this study.
If we move all 140 high-$z$ galaxies in Fig. \ref{fig-lir_td} to $z=0.015$ 
  that is a median redshift of local galaxies,
  the expected flux densities at 140 and 160 $\mu$m for more than 99\% galaxies 
  are above the minimum values of flux densities
  for our local samples ($S_{140}=1.9$ and $S_{160}=0.25$ Jy),
  which means that no galaxies would be missed with the {\it AKARI} detection limit.
This confirms that the existence of cold galaxies at high redshifts 
  that are not seen in the local sample,
  is not simply because of the different selection effect.
  
For the extreme case,
  if we move high-$z$ galaxies to $z=0.119$ that is a maximum redshift of local galaxies,
  67\% of the galaxies would be observed.
The effect of the {\it AKARI} detection limits in the $T_{\rm dust}-L_{\rm IR}$ plane is
  shown as a thick long-dashed line in Fig. \ref{fig-lir_td}.
Galaxies lying on the right of this line would be detected. 
The steep slope of the line, indicates that the {\it AKARI} detection limit 
  does not affect $T_{\rm dust}$ distribution significantly.
Moreover, cold galaxies with $T_{\rm dust}\la27$ K and $L_{\rm IR}> 3\times10^{11} {\rm L}_\odot$ 
  are free from this selection effect.
This again confirms that the existence of cold galaxies at high redshifts 
  is not because of the difference in selection effects.

To check how $T_{\rm dust}$ measurements for our local galaxies become different
  if we had flux measurements at longer wavelengths ($\lambda>160\mu$m) like for high-$z$ galaxies,
  we have re-estimated $T_{\rm dust}$ for 44 galaxies found in common between this study 
  and \citet[D00]{dun00}
  by combining our flux densities with 850 $\mu$m observations from D00.
We found that the two estimates agree very well with a median difference of 0.2 K (rms scatter of 1.8 K).
In addition, when we compare the observed 850 $\mu$m flux density in D00 with
  the expected flux density from our best-fit MBB model without 850 $\mu$m data,
  two flux densities agree well within the errors.
These also indicate that $T_{\rm dust}$ estimates with {\it AKARI} bands
  are not biased toward high $T_{\rm dust}$.
%  the difference in selection effect would not contribute significantly to
%  the difference of $T_{\rm dust}$ between the local and high-$z$ galaxies.

\subsection{Cold Galaxies at high redshifts}\label{cold}

$T_{\rm dust}$ for galaxies is a function of total SFR
  per unit dust mass, the dust emissivity, and the geometry,
  which are closely related to the global star formation efficiency (SFE).
This SFE is known to be controlled by the gas density \citep{sch59,ken98law}, 
  which is connected to the spatial extent of star-forming regions.
Therefore, the existence of {\it cold} galaxies at high redshifts %($0.3<z<1.0$)
  may imply that the spatial distribution of dust in these galaxies
  is more extended than that in local galaxies with similar $L_{\rm IR}$.
This is consistent with CO observational results for some high-$z$ galaxies
  (\citealt{tac06,iono09,dad10,tac10,gen10}; see also \citealt{kav03}).
On the other hand, the existence of {\it cold} galaxies at high redshifts
  may indicate large dust masses for these galaxies compared to local galaxies.
If we assume similar gas-to-dust ratios for local and high-$z$ galaxies
  at a given $L_{\rm IR}$,
  the detection of a large gas content in high-$z$ galaxies compared to local galaxies
  could support our results \citep{tac06,tac10,dad10}.
 
Interestingly, the {\it cold} galaxies at high redshifts ($0.3 \la z \la 1.4$) in Fig. \ref{fig-lir_td}(a),   
  fill the gap between local star-forming galaxies and high-$z$ SMGs,
  connecting the two populations (see also \citealt{gmag10bumpy}).
It is noted that previous studies on $T_{\rm dust}$ measurement of SMGs
  were based on only a few photometric data points that 
  cover only long-wavelength side of the thermal SEDs.
Therefore, it is necessary to re-estimate $T_{\rm dust}$ for SMGs 
  using {\it Herschel} data to check how the locus of SMGs 
  in the $L_{\rm IR}-T_{\rm dust}$ plane changes 
  (\citealt{bmag10}; Chanial et al. 2010, in prep.; Chapman et al. 2010, in prep.).
There are 13 SMGs in common between this study and \citet{chap05}.
Among them, we have three SMGs with $T_{\rm dust}$ measured in this study
  ($T_{\rm dust}$ for the other ten galaxies are not measured 
  because they did not fulfil the selection criteria in \S \ref{data} 
  due to the low signal-to-noise ratios in some bands),
  and show them as squares in Fig. \ref{fig-lir_td}(a).
They are found to be in the locus of known SMGs,
  indicating that our measurements are consistent with those with 
  ground-based submillimeter studies.
However, we can not determine $T_{\rm dust}$ for SMGs 
  at $3\times10^{10}{\rm L}_\odot \la L_{\rm IR}\la 7\times10^{11} {\rm L}_\odot$,
  which is important for understanding the connection between local galaxies and SMGs
  (discussed in \citealt{bmag10}).
  
For these 13 SMGs, we have checked the `clean' flag defined in this study,
  and show them in Fig. \ref{fig-lir_tdlow}: there are seven `clean' and six `blended' galaxies.
The three SMGs with $T_{\rm dust}$ values measured in this study, are found to be `clean' galaxies.
However, other three SMGs with $T_{\rm dust}<20$ K in Fig. \ref{fig-lir_tdlow}
  are found to be `blended' galaxies.
We have checked whether or not the redshifts of neighbouring sources are different from these SMGs,
  and found that the redshifts of close neighbours for two SMGs 
  (J$123636.75+621156.1$ and J$123651.76+621221.3$) 
  are not similar to their companion SMGs.
This indicates that the low $T_{\rm dust}$ of SMGs could be caused by 
  an overestimation of the FIR/(sub)millimeter fluxes due to blending problems.
One SMG (J$123721.87+621035.3$ at $z\sim0.978$) has a close neighbour at $z\sim0.969$
  that is separated by $8.6\arcsec$ ($\sim70$ kpc),
  which suggests that galaxies in an interacting pair could be colder than isolated galaxies
  (e.g., \citealt{tac06}). 
This issue -- the relation between the galaxy merging stage and $T_{\rm dust}$ --
  needs to be investigated with a large number of galaxies.
    
%You should check whether the redshifts are different or not of the neighboring 24um sources. If they are indeed different
%then insist on the bias due to flux boosting by close neighbors which could explain the coldest SMGs, hence an observational
%bias. If they are not different, then you should discuss the fact that it is puzzling that close pairs, i.e. mergers, are colder
%than apparently isolated galaxies. But this could be explained in the following way: apparently isolated galaxies are
%later stage mergers where the two galaxies cannot be isolated while pairs are ealry stages in the merger.You may
%then conclude that Tdust could be used an indicator of the merger stage of galaxies.
%Therefore, low $T_{\rm dust}$ for SMGs could be caused by 
 % the overestimation of the (sub)millimeter fluxes due to the blending problem.

\section*{Acknowledgments}
PACS has been developed by a consortium of institutes led by MPE (Germany) and including 
UVIE (Austria); KU Leuven, CSL, IMEC (Belgium); CEA, LAM (France); MPIA (Germany);
INAFIFSI/OAA/OAP/OAT, LENS, SISSA (Italy); IAC (Spain). This development has been supported by 
the funding agencies BMVIT (Austria), ESA-PRODEX (Belgium), CEA/CNES (France), DLR (Germany),
ASI/INAF (Italy), and CICYT/MCYT (Spain). 
SPIRE has been developed by a consortium of institutes led by Cardiff University (UK) and 
including Univ. Lethbridge (Canada); NAOC (China); CEA, LAM (France); IFSI, Univ. Padua (Italy); 
IAC (Spain); SNSB (Sweden); Imperial College London, RAL, UCL-MSSL, UKATC, 
Univ. Sussex (UK); and Caltech, JPL, NHSC, Univ. Colorado (USA). This development has been 
supported by national funding agencies: CSA (Canada); NAOC (China); CEA, CNES, CNRS (France); 
ASI (Italy); MCINN (Spain); Stockholm Observatory (Sweden); STFC (UK); and NASA (USA).
The HerMES data was accessed through the HeDaM database (http://hedam.oamp.fr) 
operated by CeSAM and hosted by the Laboratoire d'Astrophysique de Marseille.
This research is based on observations with AKARI, a JAXA project with the participation of ESA.
MGL was supported by the a National Research Foundation of Korea (NRF)
grant funded by the Korea Government (MEST) (grant no. R01-2007-000-20336-0).

\bibliographystyle{mn2e} % style aa.bst
%\bibliography{/Users/hhwang/Research/IDL_lib/ref_hshwang.bib} % your references Yourfile.bib
\bibliography{MNref_hshwang} % your references Yourfile.bib

\label{lastpage}

\end{document}